# Techniques for Improving the Energy Efficiency of Mobile Apps: A Taxonomy and Systematic Literature Review

Stefan Huber*, Tobias Lorey†, and Michael Felderer‡

*†‡*University of Innsbruck*, 6020 Innsbruck, Austria
‡*German Aerospace Center (DLR), Institute for Software Technology*, 51147 Cologne, Germany
‡*University of Cologne*, 50923 Cologne, Germany
ORCID: *0000-0001-7229-9740, †0000-0003-0181-1844, ‡0000-0003-3818-4442

*Abstract*—Building energy efficient software is an increasingly important task for mobile developers. However, a cumulative body of knowledge of techniques that support this goal does not exist. We conduct a systematic literature review to gather information on existing techniques that allow developers to increase energy efficiency in mobile apps. Based on a synthesis of the 91 included primary studies, we propose a taxonomy of techniques for improving the energy efficiency in mobile apps. The taxonomy includes seven main categories of techniques and serves as a collection of available methods for developers and as a reference guide for software testers when performing energy efficiency testing by the means of benchmark tests.

## I. INTRODUCTION

A serious limitation of smartphones is that they are battery-powered. This means that a low battery might prevent users from executing important tasks. Energy efficiency is therefore considered an important non-functional property of mobile applications. Given the combined energy-consumption of billions of mobile devices used worldwide, energy efficiency of mobile apps is closely related to green software engineering [1]. Together with sustainability, both are emerging topics in many disciplines and are increasingly relevant in software engineering as well. Research has shown that mobile developers are aware of energy consumption issues and are interested in solutions to these problems [2]. However, the two main obstacles which prevent mobile developers to build energy efficient software systems are the lack of knowledge and lack of tools [3].

We therefore identified the need to investigate and summarize available techniques for improving the energy efficiency of mobile apps from the literature. To address this need, we conducted a large-scale systematic literature review (SLR) that aimed to compile a representative sample of the existing body of knowledge on the topic. To enhance the accessibility and usefulness of the SLR for both practitioners and researchers, we organized our findings into a taxonomy. By navigating the categories of the taxonomy, practitioners can easily access relevant and applicable research results, while researchers can find related research that is relevant to their current research endeavors.

This paper is structured as follows: Section II provides an overview of related work on energy consumption and energy efficiency of mobile devices and applications. Section III states the research questions and explains the research methodology, containing the systematic literature review and taxonomy development approach. In Section IV, we present the results and provide a detailed description of the taxonomy. Section V discusses our findings. Section VI reflects on threats to validity and Section VII presents our conclusion.

## II. RELATED WORK

This section presents survey-oriented related work on the topic of energy efficiency of mobile devices. We identified several research articles that motivate our research process.

With the goal to identify practices that increase energy efficiency, Cruz & Abreu [4] analyzed Android and iOS apps. They organized their findings in a catalog, consisting of 22 relevant design patterns, that help guide mobile developers build more energy efficient apps. Hans et al. [5] surveyed several ways that help mobile developers create more energy efficient software. They found that especially the graphical user interface provides room for optimization, such as replacing areas displaying light colors and reducing brightness. Other improvements include using Wi-Fi for downloading instead of using cellular networks and activating the GPS sensor only in select situations.

Pasricha et al. [6] focused on mechanisms to reduce energy-consumption in both mobile and IoT devices. They surveyed the literature and provided guidance on ways to minimize energy-consumption related to hardware components, software components, battery management and user-aware mechanisms.

Marimuthu et al. [7] identified high energy-consumption of many Android applications as a problem to their developers. The authors conducted a review of 25 existing studies on automated energy diagnosis tools in the Android ecosystem. Results were used to propose a taxonomy of approaches to identify and handle energy inefficiencies containing both developer-centric as well as user-centric approaches. A systematic mapping study was conducted by Moreira et al. [8] on testing the energy efficiency of Android apps. Identified studies were selected and classified into categories. They found that research is most concerned with fine-grained estimation, test generation, and classification activities.

Dornauer & Felderer [9] conducted a comprehensive systematic literature review, including 44 primary studies aiming to provide an overview of both energy-draining as well as energy-saving aspects of mobile web apps. In addition, they investigated energy-saving experiments. They found that CPU-efficient scheduling was the most prominent technique to reduce energy consumption, followed by filtering web content and code-related optimization. Typically, these are implemented and experimentally evaluated in Android devices.

## III. METHODOLOGY

As mobile app developers lack resources in the form of knowledge and tools [3] to implement strategies that support energy efficiency. Our goal is to systematically review and summarize existing techniques to improve the energy-efficiency of mobile apps from the research literature. To ensure a well-defined scope and clear objective for this research endeavour, we applied a goal, question, metric (GQM) template [10], which is detailed in Table I.

TABLE I
GQM RESEARCH GOAL DEFINITION

| | |
|---|---|
| **Analyze** | validated techniques to improve the energy efficiency of mobile apps (or parts of mobile apps) |
| **for the purpose of** | optimizing the energy efficiency of mobile apps |
| **with respect to** | relevance of the outcome |
| **from the viewpoint of** | mobile app developer |
| **in the context of** | the primary studies that report techniques and validation thereof |

For guiding the research process the following two research questions were derived from the GQM research goal definition:

**RQ1** *What kind of techniques for improving the energy efficiency of mobile apps are reported in the research literature?*

**RQ2** *How can existing techniques for improving the energy efficiency of mobile apps be synthesized effectively to support practitioners in information retrieval and foster future research?*

In the remainder of this section we describe the research methodology in detail, which is applied to answer the research questions. Firstly, by performing a SLR in the context of energy efficiency of mobile apps and secondly, by developing a taxonomy of techniques for improving energy efficiency. Our research process closely follows existing scientific guidelines on systematic literature reviews [11], [12] and principles on developing taxonomies [13]. The followed research process is depicted in Fig. 1.

### A. Systematic Literature Review

*1) Search Strategy:* A well-defined systematic literature review is an unbiased method to analyze evidence related to the field of interest. Therefore we followed the PICO (population, intervention, control and outcome) criteria [11] for a guided definition of an appropriate search string. As our literature review is conducted to research techniques to optimize the energy efficiency of mobile apps, we focused on an appropriate definition of the population and the expected outcome of the techniques. Terms for intervention and control were omitted from the PICO criteria to have a broad and unbiased result set. Additionally, we extracted recurring synonyms for the terms by inspecting titles of well-known studies in the research area. The final terms and synonyms for the search query are listed in Table II. The full search string was generated by combining the terms and synonyms with logical operators. The resulting search string is depicted below:

> ( "mobile" OR "android" OR "ios" OR "smartphone" OR "smartphones" ) AND ( "app" OR "application" OR "applications" OR "apps" ) AND ( "energy" OR "battery" OR "power" )

We performed the literature review by applying the search string to the four databases: IEEE Xplore, ACM Digital Library, ScienceDirect, and Wiley. Before beginning the search procedure, we conducted an appropriate pretest on the search interfaces to ensure a consistent applicability of the search string on all selected databases. To ensure a comprehensive review and the possibility for replication, we selected all papers up to the end of 2022. An overview of the total results from each databases is given in Table III.

TABLE II
DEFINITION OF SEARCH QUERY

| | Term | Synonyms |
|---|---|---|
| Population | mobile | android, ios, smartphone, smartphones |
| Population | app | application, apps, applications |
| Outcome | energy | battery, power |

*2) Study Selection by Criteria:* To ensure the rigor of our SLR, we followed the guidelines set forth by [11] and established clear inclusion and exclusion criteria, based on our research question. We defined these criteria a-priori, before conducting the search process. The full set of selection criteria are listed below:

**Inclusion Criteria**

- **IC1:** Primary study, which is reporting a technique or an approach to reduce or influence the energy consumption of a mobile app.
- **IC2:** The dependent variable of the study is the energy consumption. The study is not using a proxy, such as CPU usage or execution time, to make a potential assessment of the energy consumption.
- **IC3:** Study subjects must be mobile apps or parts of mobile apps which are executable on a battery-powered mobile device or handheld device.

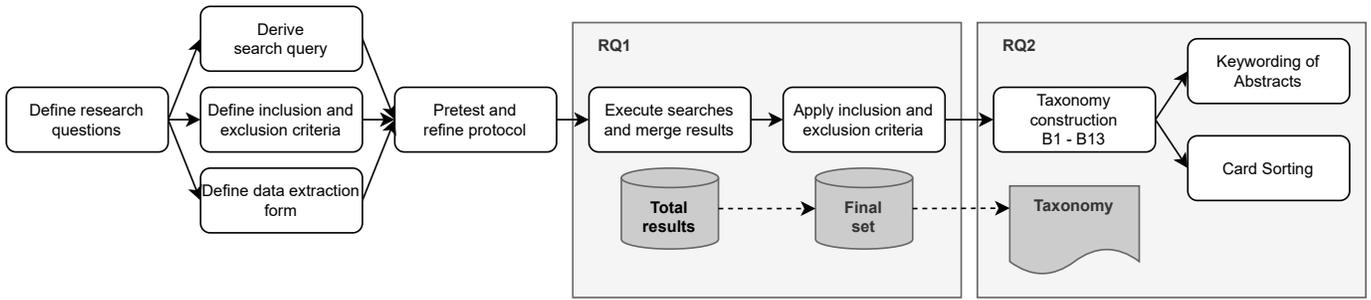

Fig. 1. Depiction of the applied research process

- **IC4:** Energy efficiency optimization techniques with reported results must be in the realm of the mobile app developer, i.e., the application level is primarily considered.

**Exclusion Criteria**
- **EC1:** Energy efficiency optimization techniques which are not directly or generally applicable by mobile app developers (e.g., hardware optimizations, operating system optimizations, low level network protocol optimization).
- **EC2:** The proposed approach was not thoroughly evaluated (e.g., only expected results are given).
- **EC3:** The paper is not a primary study (e.g., the paper is a survey paper).
- **EC4:** The paper is not reported in English language.

The title and abstract of a research paper were primarily used for applying the inclusion or exclusion criteria. When necessary also the content of the full paper was consulted, for further clarification. As an initial step, two researchers independently applied the selection criteria on the 137 papers from the ACM Digital Library database. Throughout this process, the selection criteria were clarified with an accompanying discussion, based on uncertain assignments. Also a set of decision rules was established, to ensure further consistency among the researchers, with respect to the selection procedure. The papers from the other 3 databases were split among the researchers for an independent selection. Finally, the selection procedure resulted in a set of 91 included research papers, as depicted in Table III.

TABLE III
SEARCH RESULTS BY DATABASE

| Database | Results | Excluded | Included |
|---|---|---|---|
| IEEE Xplore | 583 | 529 | 54 |
| ACM Digital Library | 137 | 111 | 26 |
| ScienceDirect | 72 | 63 | 9 |
| Wiley | 63 | 61 | 2 |
|  | 855 | 764 | 91 |

*3) Data Extraction Form:* To effectively execute the SLR process and efficiently handle the resulting papers for the taxonomy construction, a well-defined data extraction form was setup. The data extraction form plays a critical role in organizing the information gathered during the SLR process. It includes essential meta data about the publication such as the title, author names, abstract, publication year, publication type, and source database (i.e., ACM Digital Library, IEEE Xplore, ScienceDirect, Wiley), along with the DOI and a link for easy retrieval of the paper for future reference.

To aid in taxonomy construction and facilitate navigation, a comment field was included to provide a summary and a note on a potential area for classification, as well as main and sub-category fields to categorize individual papers within the taxonomy construction.

The final version of the extraction form is included in our replication package and can be utilized by practitioners and researchers alike to find relevant research papers based on the taxonomy navigation. This ensures greater accessibility and ease of use for those seeking to replicate or expand upon our SLR process.

*B. Taxonomy Development*

For summarizing the results of the SLR in an accessible and useful way, a taxonomy was constructed. To ensure the construction process was systematic and transparent, we followed the taxonomy development approach introduced by Usman et al. [13]. This approach consists of 13 steps, which we carefully followed and describe below:

**B1 - Define SE knowledge area:** Our research mainly targets the SE knowledge area Software Construction [14], as we focus on energy efficiency techniques that can be applied by and are useful for software developers of mobile apps.

**B2 - Describe the objectives of the taxonomy:** Our objective in building the taxonomy is to expand and organize the existing body of knowledge on techniques to optimize the energy-consumption in mobile apps.

**B3 - Describe the subject matter to be classified:** The taxonomy is used to classify techniques which improve energy efficiency in mobile apps.

**B4 - Select classification structure type:** We proposed a hierarchy-structured classification to fit the research question and anticipated relationships among constructs.

**B5 - Select classification procedure type:** A qualitative classification procedure was used to group and classify the energy efficiency techniques identified in the preceding literature review. First, keywords from the abstracts of the selected research papers were identified and extracted, as proposed by Petersen et al. [15]. Subsequently, the participating researchers conducted an iterative card sorting procedure [16], organizing research papers into groups based on keyword similarities. For each group an appropriate name was derived from the corresponding set of keywords. Accordingly, the card sorting approach was also applied to form the second level of the taxonomy.

**B6 - Identify the sources of information:** The sources of information used for the taxonomy construction are existing studies, which were identified throughout the execution of the proposed SLR. These studies constructed or applied the techniques for energy efficiency and validated them in experiments.

**B7 - Extract all terms:** We extracted all terms and concepts used for building the taxonomy during the course of an SLR on energy efficiency related to mobile apps.

**B8 - Perform terminology control:** We did not perform terminology control as all terms were sourced and in a following step aggregated from the literature.

**B9 - Identify and describe taxonomy dimensions:** Identified groups of techniques for increasing energy efficiency in mobile apps are selected as taxonomy dimensions.

**B10 - Identify and describe categories of each dimension:** The taxonomy contains the seven top-level categories *sensor usage, user interface, code-level optimization, application design, network usage, self-adaptation,* and *computation offloading*. A more detailed description of all the dimensions is presented in Section IV-B.

**B11 - Identify and describe the relationships:** Identified relationships are orthogonal and are organized accordingly in the taxonomy.

**B12 - Define the guidelines for using and updating the taxonomy:** We provide a detailed description of the taxonomy in Section IV, including use cases extracted from the selected research papers. We suggest further ideas for research and ideas on updating the taxonomy in Section V and Section VII.

**B13 - Validate the taxonomy:** To categorize the final set of research papers, the participating researchers performed a grouping into top-level and second-level categories. The process was guided by the card sorting technique. The grouping procedure was iterated until consensus was reached, among the participating researchers.

## IV. RESULTS

### A. Systematic Literature Review

In consideration of the first stated research question, we could identify a total of 91 research papers to be included as a result of the SLR. The first relevant paper was found in 2000, and the publication count per year is shown in Fig. 2. The peak was reached in 2015, with a decrease in publication output in recent years, in accordance with our selection criteria. The majority of papers were conference publications rather than journal articles.

After developing the taxonomy, seven top-level categories have been identified. Fig. 3 displays the distribution of publications across these categories. The category that received the highest number of publications pertains to the optimization of network usage followed by code-level optimization, while the category with the lowest number of publications focuses on application design.

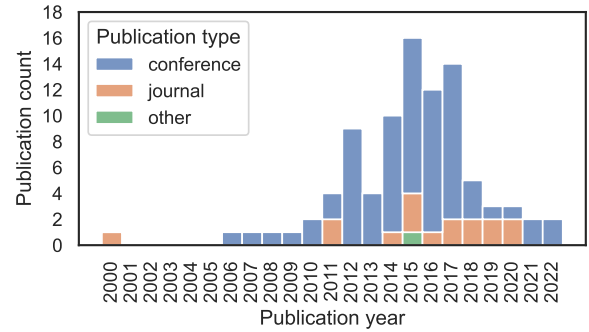

Fig. 2. Number of publications and corresponding publication type

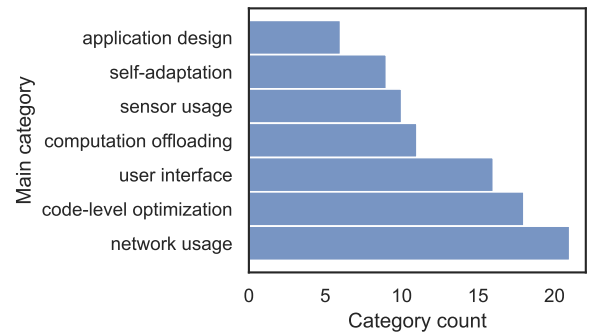

Fig. 3. Distribution of the number of papers per main category

### B. Taxonomy

Considering our second stated research question, we systematically constructed a comprehensive taxonomy to classify the existing body of research on techniques for improving the energy efficiency of mobile apps. The taxonomy is described along with its concepts and hierarchy in the following. The taxonomy contains the seven main categories *sensor usage, user interface, code-level optimization, application design, network usage, self-adaptation,* and *computation offloading*. Each of these main categories is further divided into more granular sub-categories of individual techniques. Fig. 4 gives an overview of the full taxonomy.

Techniques optimizing energy consumption based on an app's **sensor usage** often do so by targeting the *sensor selection policy* and *sensor sampling frequency*. Sensor selection policy refers to techniques that select which and how a smartphone's sensors should be used under which circumstances based on various metrics. Sensor sampling frequency techniques optimize when a sensor should be active based on defined triggers. Man et al. [17] proposed a framework that allows developers to implement solutions which activates sensor data collection based on a user's physical location. We identified one article by Lane et al. [18] suggesting piggybacking as a technique to increase energy efficiency. The idea of piggybacking it to exploit situations for collecting sensor data, in which the smartphone performs certain computations such as phone calls or sending data to the cloud.

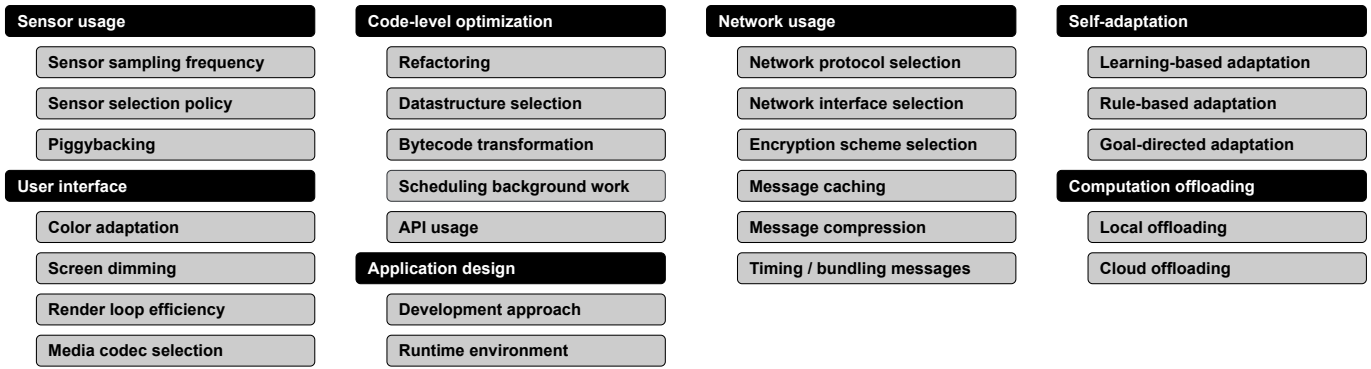

Fig. 4. Taxonomy of energy optimization techniques for mobile development

Techniques in the **user interface** category often aim at *color adaptation* used in a mobile app's GUI based on their effect on display energy consumption. Linares-Vasquesz et al. [19] proposed a solution to generate alternative energy-saving color schemes without deviating too far from the original GUI. Other techniques include *screen dimming* to reduce display brightness, *render loop efficiency*, and *media codec selection* to choose the optimal data format of audio and video files. For example, Lin et al. [20] aimed at reducing energy consumption in the context of streaming apps. They proposed an algorithm to determine optimal usage of the smartphone's built-in backlight, in the context of LCD screens.

**Code-level optimization** relates to techniques for reducing energy consumption by improving an app's code. We identified techniques in the five sub-categories *refactoring*, *data structure selection*, *bytecode transformation*, *scheduling background work*, and *API usage*. For example, Ribeiro et al. [21] developed a plugin for Android Studio to automatically refactor code to include proven energy efficient patterns. Calder & Marina [22] demonstrated that energy can be saved through batch scheduling of recurrent mobile apps.

Techniques that target the **application design** to reduce energy consumption regularly relate to the selection of the *development approach*. This includes activities such as selecting a suitable programming language, as suggested by Mahadevappa & Figueira [23]. Further, deciding between building a native mobile app or a web app to reduce energy consumption was studied by Kurtz et al. [24]. Similarly, the use of mobile cross-platform development was considered by Dorfer et.al [25]. Also the *runtime environment* can be a target for energy-saving techniques.

Techniques in the category **network usage** aim at mechanisms to enhance energy consumption related to networking tasks of the mobile app. *Network protocol selection* and *network interface selection* directly refer to core network processes, while *encryption scheme selection* improves the energy efficiency by the means of improved data encryption, e.g., during transmission. Techniques improving messaging data and processes are *message caching*, *message compression*, and *timing or bundling messages*. For example, DelBello et al. [26] suggested a strategy based on file size, data communication network, and an encryption algorithm. They posit that using Wi-FI is the most energy efficient technique for WBAN communication and that sending larger files less frequently is preferred over sending smaller files more often.

Techniques that rely on **computation offloading** distinguish between *local offloading* and *cloud offloading*. Local offloading enables applications to offload tasks e.g. between the GPU and CPU of the smartphone. Prakash et al. [27] demonstrated an approach that executes the application kernel between both the CPU and GPU. Their static partitioning approach also regulates voltage-frequency that enables energy savings. Cloud offloading techniques on the other hand approach energy saving by offloading certain computational tasks to the cloud. Oftentimes such techniques contain an algorithm to decide which tasks should be offloaded and which to execute locally. In a study by Kwon & Tilevich [28] the authors proposed the inclusion of reusable building blocks into a system architecture, that enables developers implementing cloud offloading solutions.

**Self-adaptation** refers to techniques which enable mobile apps to adapt certain characteristics to reduce energy consumption. The decision to adapt is based on either *learning-based adaptation*, *rule-based adaptation*, or *goal-directed adaptation*. For instance, Datta et al. [29] proposed a development framework that includes an analyzer engine. Based on available information including the battery level and context, one of three self-adaptive profiles is being used.

The resulting taxonomy can be described as a *Type I theory for analyzing* according to Gregor's [30] classification of theory types. Its purpose is to classify techniques that improve energy efficiency of mobile apps by allowing both practitioners and researchers to analyze the taxonomy's concepts. According to the author [30], theories of this type are especially useful to describe and analyze knowledge when little is known about about a subject. The proposed taxonomy creates awareness for energy consumption issues among app developers. It provides a comprehensive overview of sometimes little known techniques that app developers may find helpful to implement during app development. In addition, the taxonomy may enhance test quality. Software testers can refer to the taxonomy by performing targeted energy performance tests comparing implemented techniques within one app or across different apps and to identify potential energy-related defects.

## V. Discussion

We found that the large majority of identified articles in our SLR reports on techniques related to energy consumption of mobile apps in the Android ecosystem. Surprisingly, we did not find a single publication that specifically delved into energy-saving techniques for Apple's iOS operating system. A possible explanation for this matter is, that the iOS platform is restrictive on executing experiments to analyse non-functional properties, such as the energy-efficiency. As such, researchers face significant obstacles when trying to replicate energy consumption reduction techniques for iOS. To effectively develop energy optimization techniques that can be universally applied, it is essential to establish appropriate replication mechanisms also for the iOS platform. Thus, an important avenue for future research would be to develop appropriate replication mechanisms for the iOS platform. This could for example be in the form of a simulation testbed, which mimics the platform's runtime environment as much as possible, to bypass restrictions.

A further interesting observation is that negative results are rarely published. We only came across a single publication that explicitly reports on results labeled as not reliable. In their research on the effects of adjusting deep parameters to enhance the energy efficiency of mobile applications, the authors of the study [31] clearly state that it is safe for developers to ignore this technique. We contend that such findings hold just as much significance for the research community and practitioners alike. They offer valuable insights into the specific scenarios and contexts in which a particular approach may fail to yield any significant energy savings.

Upon analyzing the distribution of identified research papers over the years, it is evident that there has been a significant decline in research output since its peak in 2015, aligning with our SLR results as depicted in Fig. 2. Despite this trend, the importance of the field remains indisputably high, given the considerable challenges in advancing battery technology [32]. Furthermore, when considering the perspective of mobile app users, it becomes clear that energy efficiency is a critical factor in enhancing their satisfaction with an app [33].

Finally, only few articles relate the energy efficiency of mobile apps to sustainability. Given the current trend of sustainability research in software engineering [34], [35], we expected a lot more articles directly referring to the aspect of environmental sustainability when investigating energy consumption.

## VI. Threats to Validity

We identified several threats to validity, which should be considered in the context of our results.

Firstly, we addressed the potential threat that the resulting research papers are not representative of the construct under study. The selection of keywords clearly impacts the number and relevance of the search results. We, therefore, selected a broad range of keywords including synonyms using the established PICO guideline [11]. A lower precision of the search results was accepted and therefore more manual inspection to mitigate potential incompleteness was taken into consideration. Furthermore, we selected four different search engines to apply our search string, resulting in a total of 855 search results.

Secondly, we addressed the subjectivity of the paper selection process for our final set. To mitigate this, we developed a set of clear inclusion and exclusion criteria before starting the process. Two of the authors independently reviewed subsets of the articles and cross-validated their findings. This process led to a refinement on the understanding of our criteria. Additionally, the authors established a set of decision rules through an accompanying discussion. These rules provided a more replicable protocol for the selection process, enhancing the validity of our final set.

Finally, to ensure that the constructed taxonomy is appropriate and useful, it's crucial to group the papers into relevant categories and use appropriate terminology for each category. To achieve this, we followed the process of extracting keywords from the abstract and then use iterative card sorting to establish the taxonomy. Throughout the process, the participating authors engaged in continuous alignment on the terminology and validation to ensure the accuracy and relevance of the taxonomy. By following this process, we ensured that the resulting taxonomy is comprehensive and effective in categorizing the existing body of research.

## VII. Conclusion

In this paper we investigated techniques that improve energy efficiency in mobile apps. We conducted a systematic literature review in four scientific databases. Based on our findings, we contributed to the body of knowledge in software engineering by systematically developing a taxonomy of techniques to increase energy efficiency of mobile apps. The presented taxonomy is useful for both research and industry. Researchers may use the taxonomy to further investigate the relationships between energy-consumption and development practices. For developers, the taxonomy serves as a collection of available approaches and evaluated best practices to improve energy-consumption when building mobile apps. We believe the taxonomy is also highly relevant in software testing. Software testers can use the taxonomy as a reference guide when performing benchmark tests e.g., in performance testing or energy efficiency testing. In the future, identified techniques part of the proposed taxonomy can be analyzed on a quantitative basis and newly discovered techniques can be reflected in the taxonomy as well.

## Data Availability

We make the data available in the form of a replication package within a public Github repository (https://github.com/stefanhuber/SEAA-2023). It includes the complete list of identified primary studies as well as decision rules on the categorization of techniques.